\documentclass[useAMS,usenatbib,usegraphicx]{mn2e}
\usepackage{amssymb,amsmath}
\usepackage{subfigure}
\usepackage{color}
\def \arcmin{\hbox{$^\prime$}}
\def \arcsec{\hbox{$^{\prime\prime}$}}
\def \ccdproc{\hbox{\sevensize CCDPROC}}
\def \ciao{\hbox{\sevensize CIAO4.1}}

\def \crrej{\hbox{\sevensize CRREJ}}
\def \daophot{\hbox{\sevensize DAOPHOT}}
\def \degr{\hbox{$^\circ$}}

\def \iraf{\hbox{\sevensize IRAF}}
\def \isis{\hbox{\sevensize ISIS}}
\def \kpnoslit{\hbox{\sevensize KPNOSLIT}}
\def \lacosmic{\hbox{\sevensize L.A.COSMIC}}
\def \mkacisrmf{\hbox{\sevensize MKACISRMF}}
\def \mkarf{\hbox{\sevensize MKARF}}
\def \msun{\hbox{M$_{\odot}$}}
\def \phot{\hbox{\sevensize PHOT}}
\def \psextract{\hbox{\sevensize PSEXTRACT}}
\def \stsdas{\hbox{\sevensize STSDAS}}
\def \synphot{\hbox{\sevensize SYNPHOT}}
\def \wavdetect{\hbox{\sevensize WAVDETECT}}
\def \xspec{\hbox{\sevensize XSPEC}}

\title[Globular cluster CVs: M22 CV1 and M5 V101]{Ground and space-based study of two globular cluster CVs: M22 CV1 and M5 V101}
\author[A.~P. Hourihane et al.]{A.~P. Hourihane$^{1}$\thanks{E-mail: a.hourihane@ucc.ie}, P.~J. Callanan$^{1}$, A.~M. Cool$^{2}$ and M.~T. Reynolds$^3$\\
$^{1}$Department of Physics, University College Cork, Cork, Ireland\\
$^{2}$Department of Physics and Astronomy, San Francisco State University, 1600 Holloway Avenue, San Francisco, CA 94132, USA\\
$^3$Department of Astronomy, University of Michigan, 500 Church Street, Ann Arbor, MI 48109, USA}
\begin{document}


\pagerange{\pageref{firstpage}--\pageref{lastpage}} \pubyear{2002}

\maketitle

\label{firstpage}

\begin{abstract}
As a class of compact binaries with large binding energy, cataclysmic variables formed through close encounters play an important role in the dynamical evolution of globular clusters. As part of a systematic search for CVs undergoing dwarf nova eruptions in globular clusters, our 2004 monitoring programme of M22 detected an outburst of the dwarf nova candidate CV1 during May. We implement the \isis{} image subtraction routine to obtain a light curve for an outburst of CV1. We present the outburst light curve as well as \textit{HST}/WFPC2 photometry in the \textit{V}, \textit{U} and near ultra-violet (\textit{nUV}) bands and a \textit{Chandra}/ACIS-S spectrum of the object. Our results confirm the DN nature of the outburst and the CV status of the object. We also present the results of a ground-based study of another globular cluster CV, M5 V101 - including quiescent medium-resolution WHT/ISIS spectroscopy in the {\it B} and {\it R} bands, displaying prominent Balmer and He\,{\sevensize I} emission, and {\it R}-band photometry. 
\end{abstract}


\section{Introduction}
Compact binary systems in globular clusters (GCs) are an important target for observations. Many of these systems, unlike primordial field binaries, are thought to be formed dynamically -- for example, via three-body encounters and exchange interactions \cite[e.g.][]{hut83,ivanova06} -- often leading to very close binaries, possibly with different properties relative to their counterparts in the field. Cataclysmic variables (CVs) are compact binaries comprising a white dwarf primary accreting from a low-mass Roche-lobe-filling secondary companion, usually through an accretion disc. An increase in mass-transfer through the disc can lead to a brightening of the system of several magnitudes, known as a dwarf nova (DN) eruption. 

So far, relatively few DN outbursts of globular cluster CVs have been observed \cite[e.g.][]{shara96,tuair03,piet08} compared to the predicted populations of such systems. \cite{ivanova06} predict from numerical simulations that the number of CVs per unit mass should be 2--3 times \emph{higher} in GC cores than in the field.
\citet{piet08} estimate the number of erupting CVs per unit mass in the solar neighbourhood to be $1.9\text{--}4.8\times10^{-5}$ per \msun. However, they find that the number of CVs per unit mass in GCs which have been observed to undergo DN eruptions is significantly ($\sim10$ times) \emph{smaller} than in the local field. 
It is likely that selection effects play a role in this apparent discrepancy. For example, \citet{gansicke05} notes that out of 531 field CVs with known orbital periods, those systems discovered through their variability are comprised of 86 per cent classical and dwarf novae, whereas more than one half of the 121 X-ray selected systems in the sample are magnetic systems, which erupt relatively infrequently, if at all.
On the other hand it is 
plausible that despite the role of selection effects, the observational evidence for a lack of outbursts in GC CVs nevertheless reflects an underlying difference in the nature of the cluster CVs compared to field CVs. 

As well as the bright GC X-ray sources ($L_{\mathrm{X}} \gtrsim
10^{36}$~erg~s$^{-1}$) which have been identified with low-mass X-ray
binaries \citep[LMXBs;][]{grindlay84}, a population of fainter
($L_{\mathrm{X}} \lesssim 10^{33}$~erg~s$^{-1}$) X-ray sources in a
number of GCs has been more recently identified as being composed of a
significant proportion of CVs. Using the {\it Hubble Space Telescope}
(\textit{HST}), \citet{cool95} optically identified the first sample
of CVs in a GC, coinciding with a population of \textit{ROSAT} X-ray
sources in NGC6397 \citep{cool93} -- the current number of CV
identifications in the cluster is $\sim15$ \citep{cohn10}. Meanwhile,
the largest sample of CVs (22) in a GC has been identified in 47 Tuc
\citep[for a list of the observations, see][]{heinke05}. 
This number compares favourably (within a factor of 2) to the theoretical prediction of \citet{ivanova06} of 35--40 \emph{detectable} CVs (and up to 200 total CVs) in a 47 Tuc-like GC (where observational limits for field CVs are imposed to define detectability). The
confirmed CV populations in these and other GCs allow us to compare
the properties of these predominantly dynamically-formed GC CVs to
those of primordial field CVs and also to theoretical models.

It appears there is some evidence for a CV population in GCs with different characteristics to the field population. 
Many GC CVs have higher X-ray luminosities than field CVs \citep{verbunt97}, for instance. Many also exhibit X-ray to optical flux ratios higher than those typically seen in field CVs \citep[this was found in the case of the 47 Tuc CVs by][]{edmonds03b}, indicative of dwarf novae -- but in contradiction to the aforementioned dearth of observed DN eruptions. Furthermore, the presence of He\,{\sevensize II} emission lines in the {\it HST\/} spectra of some GC CVs suggests a possible magnetic nature \citep[as proposed for CVs in NGC6397 by][]{grindlay95}. 

These globular cluster CVs are difficult to observe from the ground in their quiescent state due to their intrinsic faintness and their crowded location in the densely populated cluster cores. GC CV candidates have thus far typically been identified on the basis of their X-ray or far-UV emission (with optical follow-up observations needed for confirmation) and, in fewer instances, the presence of the more definitive DN outbursts. While the resolving power of {\it HST} is needed for the detailed optical photometric or spectral analysis necessary for conclusive classication of these crowded sources, the advent of image-subtraction software such as \isis{} \citep{alard98,alard00} allows us to search for eruptions of these systems from the ground. 

CV1 is located in the crowded central region of the nearby \cite[3.2~kpc;][]{harris96} globular cluster, M22. Previously, \cite{sahu01} interpreted a $\sim$3~mag brightening of the source in 1999 May as gravitational microlensing of a background bulge star by a low-mass foreground cluster object. Subsequent analysis of {\it HST\/} archival data by \citet*[]{ack03}, including proper-motion studies confirming the object as a cluster member, prompted them to reclassify the object as a cluster CV which had undergone a dwarf nova eruption during the observations by \citet{sahu01}. They also found that, in optical colours, the star is unusually red for a CV, with ($B-R$) and ($V-I$) colours about 0.2 and 0.1~mag redward of the main sequence, respectively. \citet{bond05} reported two outbursts of the object occurring during 2002--2003 from ground-based {\it I}-band observations during that epoch. The outburst amplitudes were $\Delta{\it I}=2\mbox{--}3$~mag and the full width at half-maximum of the duration (measurable for one outburst only) was approximately 14~d. In another ground-based study, \citet{piet05} reported a further two outbursts of CV1 -- one occurring in 2000 August/September, with a duration of $\sim20$~d and the other event, for which they had incomplete data coverage, in 2001 June. More recently \cite{webb04} have suggested that CV1 may be not be a CV, but could instead be a quiescent low-mass X-ray binary (LMXB) that exhibits outbursts. We studied M22 CV1 with the aim of placing a new constraint on the outburst frequency as well as further exploring the broadband optical colours and investigating the X-ray properties to see if a quiescent LMXB (qLMXB) identification could be ruled out for the source. We report on a long-term ground-based optical monitoring programme of M22 during which an outburst of CV1 was detected and we also present archival \textit{HST} photometry and analysis of the X-ray properties of the object.
 
Although the dwarf nova V101 lies at the greater distance of 7.5~kpc from us in the globular cluster M5 \citep{harris96}, its atypical location ten core radii from the centre of the cluster and lack of close neighbours allow it to be studied spectroscopically from the ground. A review of the available literature on V101 was given by \citet{neill02}. From ground-based {\it I}-band observations they determined the orbital period of the system to be P$_{\mathrm{orb}}=5.796\pm0.036$~h -- the first orbital period determination for a dwarf nova in a globular cluster. The only literature on the source since that time is the report by \citet{piet08} of the detection of two new outbursts in 2003 and 2004 as part of their Cluster AgeS Experiment (CASE), as well as two previous outbursts in 1997, which had already been reported by \citet{kaluz99} in their ground-based study of the cluster. We present an optical multiwavelength study of the M5 dwarf nova, V101, in an attempt to confirm its orbital period and identify the spectral type of the secondary.

\section[]{Observations and data analysis}
 
\subsection{M22: optical photometry and X-ray spectroscopy}

\subsubsection{Ground-based optical photometry}
The M22 ground-based dataset consists of {\it V} and {\it B}-band images covering the period of 2004 March--November, taken with the Small and Moderate Aperture Research Telescope System (SMARTS) 1.3-m telescope at the Cerro Tololo Inter-American Observatory (CTIO), Chile. The $1024\times1024$~pixel ANDICAM detector used for the observations has a pixel scale of 0.369~arcsec~pixel$^{-1}$ and a $6\times6$~arcmin$^2$ field of view. The images are centred on the core of the globular cluster with a sampling of approximately two 300-s {\it V}-band exposures every second night for most of the observing programme, increasing during May to two exposures per night in both the {\it V} and {\it B} bands as soon as the outbursting object was discovered.

\begin{figure}
\centering
\includegraphics[width=0.7\linewidth]{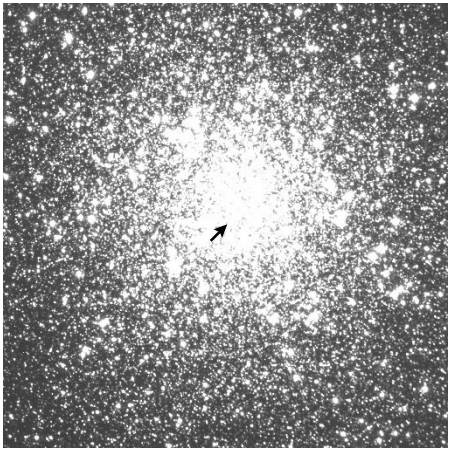}
 \caption{One of our SMARTS {\it V}-band images of M22 showing the location of CV1 near the centre of the cluster. The image clearly illustrates the crowded nature of this field.}\label{fig:m22full}
\end{figure}

\begin{figure}
\centering
\begin{tabular}{c}
\includegraphics[width=0.5\linewidth,angle=0]{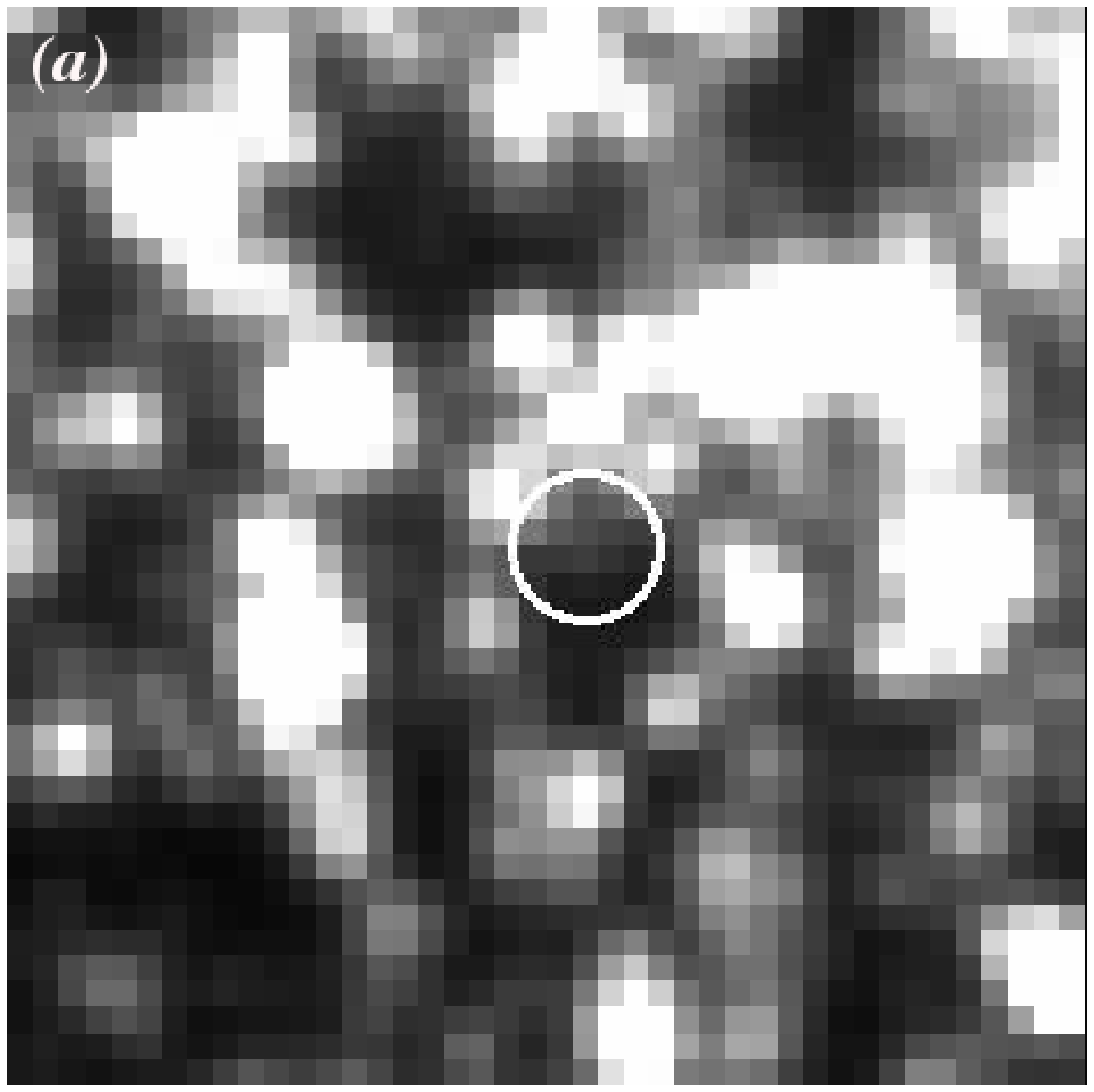}\label{fig:smarts1}
	\\
\includegraphics[width=0.5\linewidth,angle=0]{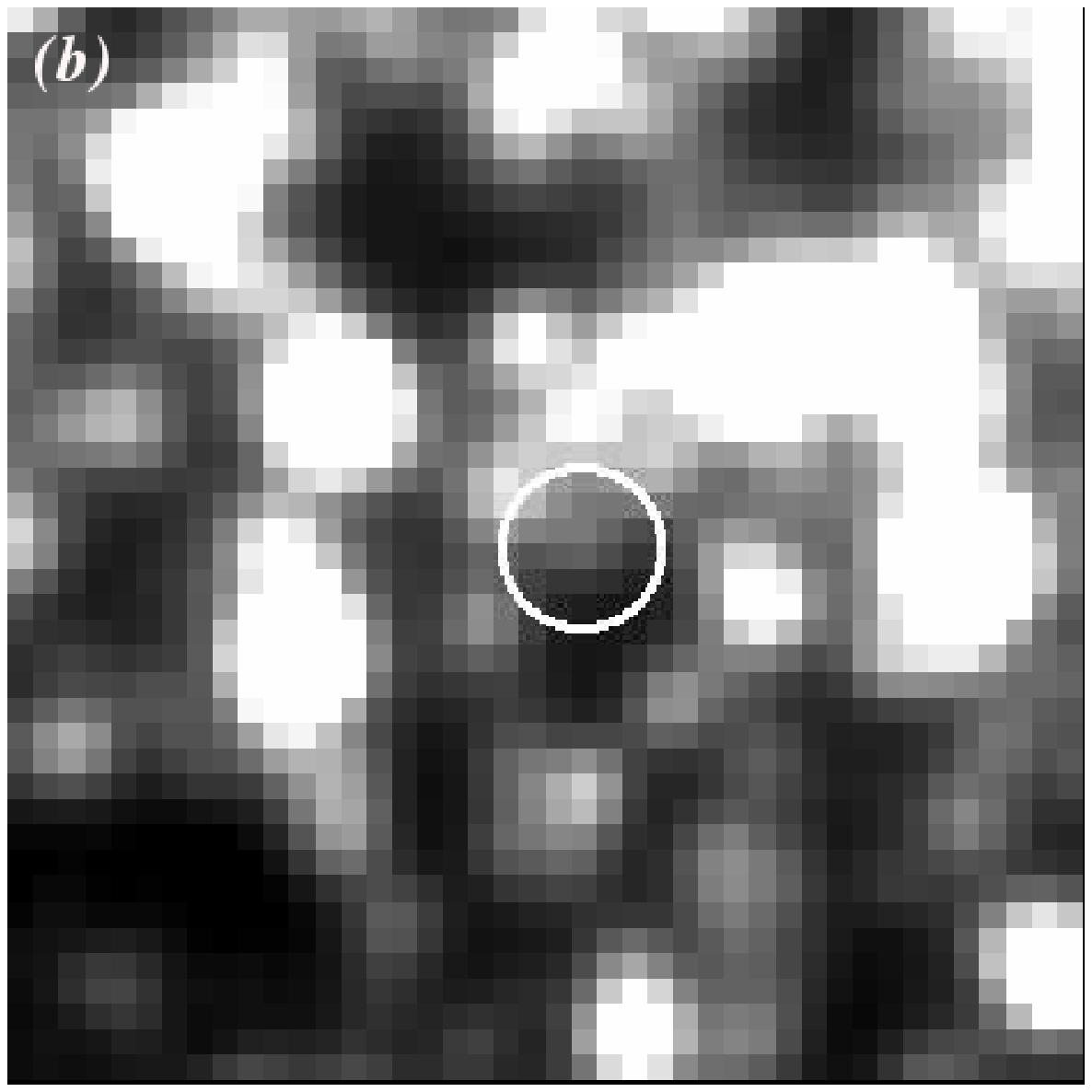}\label{fig:smarts2}
	\\
\includegraphics[width=0.5\linewidth,angle=0]{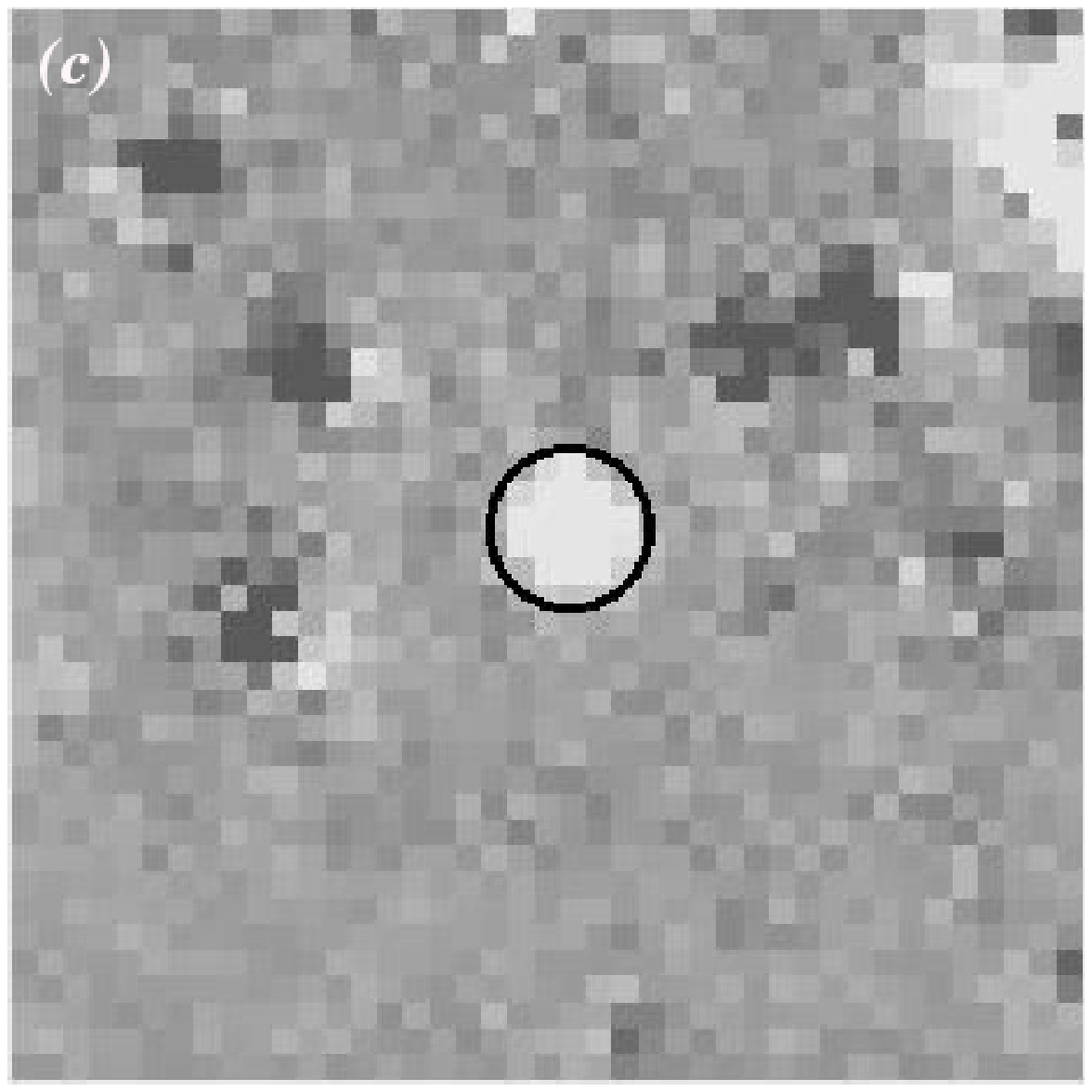}\label{fig:m22sub}
\end{tabular}
\caption{SMARTS {\it V}-band image sections centred on M22 CV1. (a) An image from the night of 2004 April 17 shows the location of the quiescent CV; (b) shows the CV in outburst on 2004 May 13. In (c), CV1 is clearly visible in outburst during 2004 May in a subtracted image processed with the \isis{} software. Residuals from saturated neighbouring stars also appear in this image.}
\label{fig:m22smarts}
\end{figure}

Due to severe crowding effects at the cluster core and the faintness of the CV (see Figs.~\ref{fig:m22full} and \ref{fig:m22smarts}), standard photometric reduction techniques such as aperture and profile-fitting photometry could not be performed on the source. For example, even small aperture photometry would be contaminated by light from neighbouring stars. Also posing a problem was the uncertain level of the sky background, which can vary quite strongly in such crowded regions due to, for one, faint unresolved background stars. Moreover, the crowded nature of the entire image meant it was impossible to construct a model point spread function of sufficient quality for profile-fitting photometry. The photometry was instead performed using the \isis{} image-subtraction software \citep{alard98,alard00}. Using the programs in this package, the images were first aligned with each other by remapping on to a common grid. A reference image for subtraction was created by combining the ten per cent of the images with the best seeing. This reference image was then convolved with a spatially varying kernel to degrade its seeing to match the seeing of each individual image in turn. Each individual image was then subtracted from the appropriately convolved reference image, with the result of creating a subtracted image containing the residuals (the outbursting CV is obvious in these subtracted images -- see Fig.~\ref{fig:m22smarts}~(c)). All the subtracted images were median-combined to produce an image, {\it var.fits}, showing -- in theory -- only the variable stars, which appear as stellar-like profiles. In practice, saturated stars and cosmic rays lead to many spurious detections. The light curve of the CV was extracted from the subtracted images in the {\it V} and {\it B} bands using the \phot{} routine packaged with \isis{} and plotted in units of differential flux relative to the reference image in which the CV was in quiescence.

\begin{table*}
 \centering
 \begin{minipage}{140mm}
  \caption{List of archival space-mission data used in the analysis}\label{table:obs}
  \begin{tabular}{@{}llllr@{}}	
  \hline
  Satellite & Instrument & ID & Date & Exp. Time \\ 	
   & & & & s \\
 \hline
 {\it HST\/} & WFPC2/F555W & U9BM0704M & 2006 April 01 & 30 \\
 {\it HST\/} & WFPC2/F336W & U9BM0703M & 2006 April 01 & 500 \\
 {\it HST\/} & WFPC2/F255W & U9BM0701M-702M & 2006 March 31 & 800 \\
 {\it Chandra \/} & ACIS-S & 5437 & 2005 May 24 & 16020 \\
 \hline
\end{tabular}
\end{minipage}
\end{table*}

\subsubsection{HST data}\label{sec:hstdata}
\begin{figure}
\centering
\includegraphics[width=0.7\linewidth]{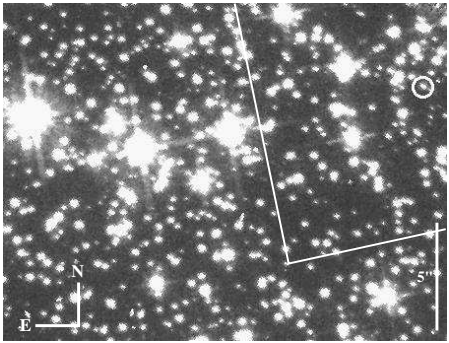}
\caption{Finding chart for M22 CV1, taken from archival {\it HST} image u9bm0704m ({\it F555W}), showing the region of overlap with the SMARTS images in Fig.~\ref{fig:m22smarts}. The quiescent CV appears at the edge of chip WF4 of the WFPC2.}\label{fig:m22find}
\end{figure}

We retrieved archival multicolour {\it HST\/} images of M22 from the Multimission Archive at STScI.\footnote{Based on observations made with the NASA/ESA Hubble Space Telescope, obtained from the data archive at the Space Telescope Science Institute. STScI is operated by the Association of Universities for Research in Astronomy, Inc. under NASA contract NAS 5-26555.} A list of the observations is given in Table~\ref{table:obs}. CV1 appears on chip WF4 of the Wide-Field Planetary Camera 2 (WFPC2), as seen in Fig.~\ref{fig:m22find}: its coordinates are $\alpha=  18^{\mathrm{h}}36^{\mathrm{m}}24^{\mathrm{s}}\!.66$, $\delta=  -23\degr54\arcmin35\arcsec\!.5$ \citep[J2000;][]{sahu01}. In order to correct the counts in each pixel of the pipeline-calibrated images for the effects of geometric distortion (which appears at the 1--2 per cent level near the edges of the images and up to a maximum of 4--5 per cent in the corners, for fixed-aperture photometry), we multiplied each image by the correction image obtained from the archive. Cosmic rays were removed in \iraf{}\footnote{IRAF is distributed by the National Optical Astronomy Observatories, which are operated by the Association of Universities for Research in Astronomy, Inc., under cooperative agreement with the National Science Foundation.} with the \stsdas{} \crrej{} tool for the co-aligned images taken with the {\it F255W} filter ($nUV_{255}$-band) and with the \lacosmic{} Laplacian edge-detection routine \citep{dokkum01} for the individual images in the {\it F336W} and {\it F555W} filters ($U_{336}$- and $V_{555}$-band, respectively). We performed small-aperture photometry in the $U_{336}$ and $V_{555}$ bands with the \daophot{} \citep{stetson87} package in \iraf{} on all objects detected -- including CV1 -- in the $V_{555}$-band image on the PC, WF2 and WF4 chips. We corrected the results to the aperture for which the zeropoint was defined. The magnitudes were calibrated on the STMAG flux-based system using the zeropoint calculated from the \synphot{} PHOTFLAM header keyword. All the magnitudes were corrected for charge transfer efficiency (CTE) losses using the correction formulae of Andy Dolphin\footnote{http://purcell.as.arizona.edu/wfpc2\_calib/} and the UV measurements were also corrected for the reduction in throughput due to the build-up of contaminants on the UV filters in between the regular decontaminations.

For CV1, we also calculated the STMAG $nUV_{255}$ magnitude and the ($nUV_{255}-U_{336}$) colour. The results for CV1 quoted 
in what follows include a correction for the red leak of the UV filters. This was estimated using \synphot{}, which is part of the \stsdas{} package in \iraf{}, combined with the newer filter-throughput data in ISR09-07\footnote{http://www.stsci.edu/hst/wfpc2/documents/isr/wfpc2\_isr0907.html}. For the purposes of comparison with catalogued values we calculated the Johnson $(U-V)_{0}$ colours for the CV using the transformation in equation 8 and the zeropoints in Table 7 of \citet{holtz95}.

\begin{figure}
\centering
\includegraphics[width=0.7\linewidth]{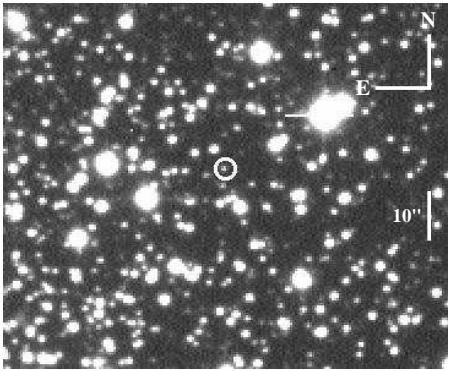}
 \caption{Finding chart for V101 in M5, taken from an archival INT {\it R}-band image.}\label{fig:m5find}
\end{figure}

\subsubsection{X-ray observations}
Our X-ray dataset consists of 16~ks of archival {\it Chandra\/} ACIS-S data, obtained from the {\it Chandra Data Archive\/} (see Table~\ref{table:obs}). The central region of the cluster where CV1 resides appears on the back-illuminated S3 chip of the ACIS detector, close to the aimpoint of the telescope. Beginning with the level-1 pipeline-processed event lists, the data were reduced with the \ciao{} software to apply the latest calibration. The data were filtered on the energy range for which the ACIS detector is calibrated, 0.3--10~keV. Bad pixels and times of high background were removed using the pipeline-produced badpixel file and good-time intervals (GTI). Source-detection was carried out with the \wavdetect{} algorithm, which correlates the data with a `Mexican hat' wavelet function. A source appearing at $\alpha=  18^{\mathrm{h}}36^{\mathrm{m}}24^{\mathrm{s}}\!.71$, $\delta=  -23\degr54\arcmin35\arcsec\!.6$ is consistent with M22 CV1: this position corresponds to a radial offset of $\sim$0.7\arcsec{} from the \textit{HST} position (see Section~\ref{sec:hstdata}), which is within the uncertainties of the \textit{HST} ($\mathbf{\sim 0.5}$\arcsec) and \textit{Chandra} ($\mathbf{0.6}$\arcsec) coordinates. We extracted the source spectrum with the \ciao{}-contributed script for point sources, \psextract. The detector response and effective-area files were extracted with the \mkacisrmf{} and \mkarf{} tools and binned on the same energy grid. A background spectrum was extracted from source-free regions neighbouring the object. The same instrument response files were used for the background as for the source. The source spectrum was grouped to include at least twenty net counts per bin. Fitting of the source and background spectra was performed in \xspec{} version 12.4.\footnote{http://heasarc.gsfc.nasa.gov/docs/xanadu/xspec/}

\subsection{M5: optical photometry and spectroscopy}
The optical spectroscopic observations of M5 were taken during 1991 July and August with the 4.2-m William Herschel Telescope (WHT) on La Palma, using both the red and blue arms of the Intermediate dispersion Spectrograph and Imaging System (ISIS). The R316R grating was used with the red arm and the R300B grating with the blue arm, producing a resolution of 3.3~\AA. The seeing during the spectroscopic observations varied from 0.7 to 1.3~arcsec. The finding chart for the CV appears in Fig.~\ref{fig:m5find}.

We used two sets of imaging data. The first consists of {\it R}-band images obtained from the ING archive.\footnote{This paper makes use of data obtained from the Isaac Newton Group Archive which is maintained as part of the CASU Astronomical Data Centre at the Institute of Astronomy, Cambridge.} The observations were taken with the 2.5-m Isaac Newton Telescope (INT) on La Palma over several nights in 1990 April and June. The uncertainty in the orbital period leads to a possible phase error of ten per cent after four days. On combining data with a smaller temporal separation than this, the resulting datasets have approximately eighty per cent orbital phase coverage. 

The second set of imaging data we obtained at the 2.4-m Hiltner telescope at Michigan--Dartmouth--MIT (MDM) observatory using a Sloan Digital Sky Survey (SDSS) {\it r}-band filter. It consists of twenty-seven 60-s exposures and seventy-eight 180-s exposures taken over four nights during 2009 July. The field of view was 4.97$\times$3.32~arcmin$^2$ while the images had a typical seeing of about 1.6~arcsec. This dataset had approximately fifty per cent orbital phase coverage.
    
All images were prepared by de-biasing, trimming and flat-fielding with the \ccdproc{} routine in \iraf. For the spectral images, an aperture was selected around the target on each image and the spectrum was extracted using routines in the \kpnoslit{} package in \iraf. The wavelength-calibration arc-lamp spectra (CuNe for the red and CuAr for the blue) were extracted using the same parameters as for the corresponding target spectra. The arc spectra were fitted with a high-order cubic spline to remove the shape of the continuum of the illuminating arc-lamp. The wavelength solutions for the target spectra were calculated using the lines identified in the arc spectra and checked using the sky lines in a sky spectrum extracted from a source-free region of the spectral image containing the target. With only three blue spectra for which we had enough signal-to-noise (S/N) to obtain a trace for extraction on the spectral image, we summed the spectra and removed cosmic rays (which were located on the continuum) manually in \iraf. We also summed the red spectra and again removed cosmic rays manually, checking for consistency against the median-combined cosmic-ray-filtered spectrum. Finally, we measured equivalent widths of any emission lines in the summed spectrum by fitting with a Gaussian line profile in \iraf, excluding from the fits data from any spectrum which had been affected by a cosmic ray hit on that emission line.  

For the imaging data, profile-fitting photometry was performed with the \daophot{} \citep{stetson87} package in \iraf. A set of bright, isolated stars is required to model the stellar profiles arising from the point spread function (PSF) of the light in an image. In a crowded field where no well-isolated stars are available, an iterative procedure is required to build the PSF. The following is the procedure we used to construct the PSF of our images. First, a Gaussian function was fitted to a sample of reasonably uncrowded stars with good signal-to-noise and no defects, in an attempt to model the PSF. A constant PSF model was initially computed and fitted to the chosen PSF stars and their neighbours in order to subtract them from the image, revealing in the process previously invisible faint neighbouring stars -- these were subsequently added to the star list, had photometry performed on them and were finally subtracted from the image. The process was iterated upon until a fairly complete list of PSF-star neighbours had been found and these were then subtracted from the original image to produce an image containing the PSF stars but excluding their close neighbours. From this image a PSF model which varied linearly across the image was computed. This new model was used to fit and subtract (more cleanly) the PSF stars and their neighbours from the original image in the same way as before, iterating on the process to eliminate the close neighbours of the PSF stars and finally produce an image containing only the PSF stars, which at this stage had become well-isolated enough to compute a sufficiently accurate model of the PSF to use for photometry. It was from this image that the final PSF model was made. This model was then fitted to the source (the CV, M5 V101) and several comparison stars to obtain their instrumental magnitudes. The process was repeated for each image and a light curve was constructed using the same brighter comparison star in each image (with constancy checked against several other comparison stars). The light curves were folded on the orbital period and plotted against orbital phase. Phase zero is arbitrary since the ephemeris was not known.

\section{M22 CV1: Outbusts and Optical/X-ray Colours}

\subsection{Optical variability}

\begin{figure*}
\includegraphics[width=0.7\linewidth,angle=270]{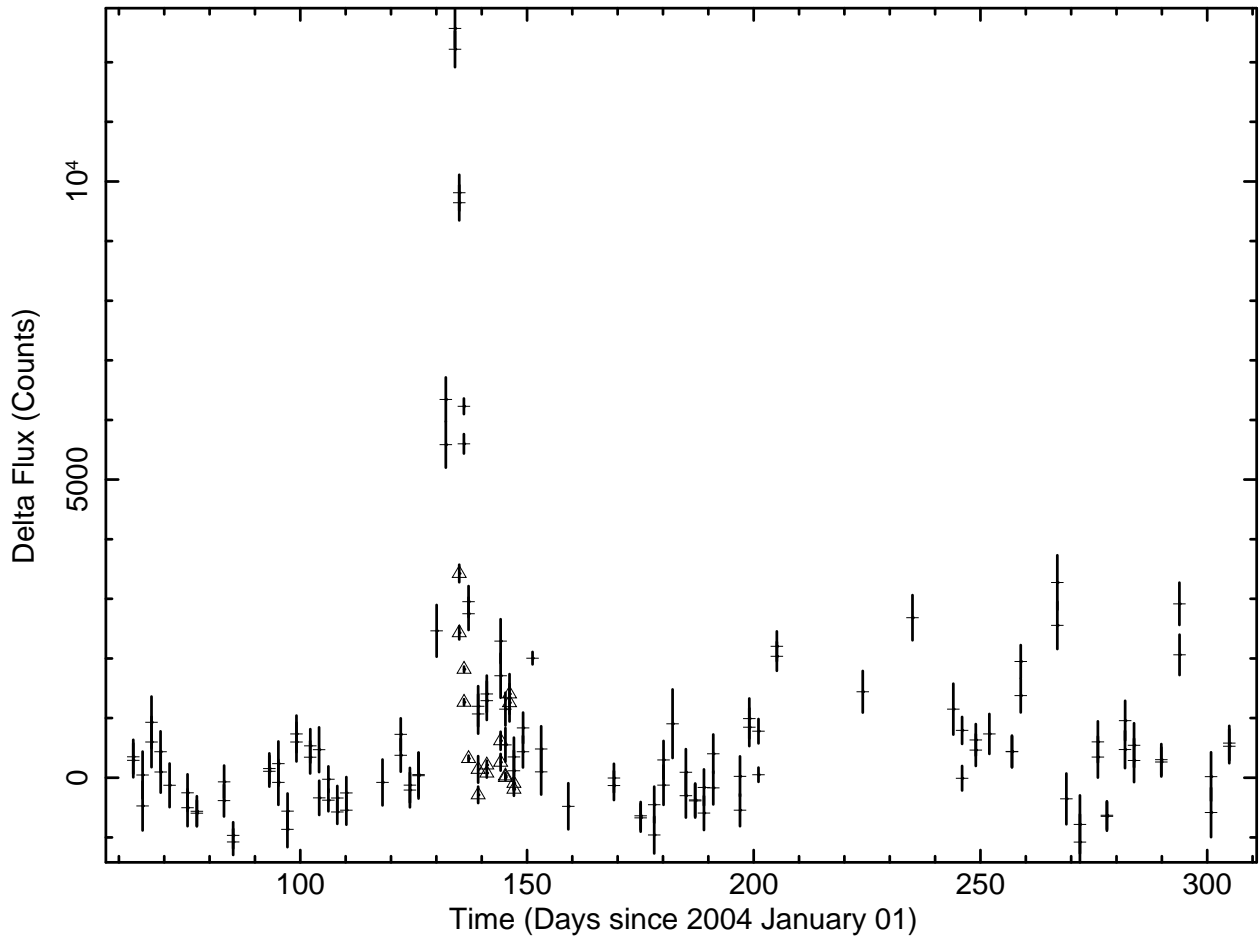}
 \caption{The \isis{} differential flux light curve showing the 2004 May outburst. The crosses represent the {\it V}-band data and the triangles the {\it B}-band data.}\label{fig:flux}
\end{figure*}

Our \isis{} light curve (Fig.~\ref{fig:flux}) shows a $\sim$15-d outburst in the {\it V} band during 2004 May, the only large outburst seen in nine months of monitoring. The rise to maximum takes place over at least five days, fading over another ten days, approximately. The actual peak of the outburst may have been missed in the $\sim$24-hr sampling gaps. We were unable to calibrate the outburst amplitude on the magnitude scale due to severe crowding effects combined with the apparent faintness of the CV. The outburst was also caught in the {\it B}-band; however, due to the aforementioned difficulty in calibrating the flux, it is unclear if the lower {\it B}-band flux-levels compared to the {\it V}-band are real or not. The best available reference image to use for the {\it B}-band reductions came from the end of May when the flux level had more or less reached quiescent levels again -- however, there may still have been some activity of the object at this time, which would, in effect, reduce the flux measured in the subtracted images. Inspecting the full dataset, we see low-level flickering-like variability, typical of CVs, which appears to be present at a level above the noise -- but again, caution is required in interpreting this feature without proper calibration.

\begin{figure}
\includegraphics[width=0.7\linewidth,angle=270]{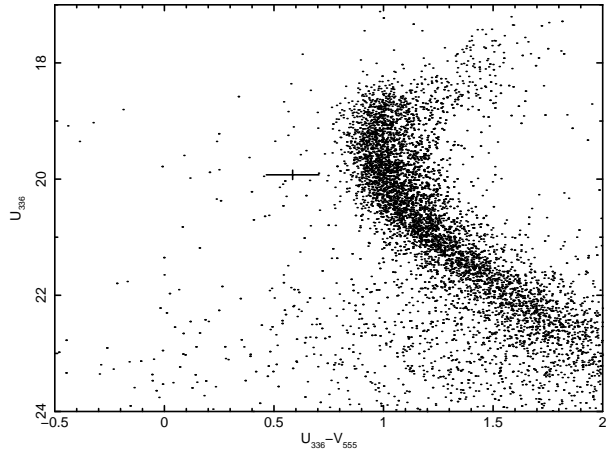}
 \caption{$U_{336}$ versus ($U_{336}-V_{555}$) colour-magnitude diagram for M22, with $U_{336}$ and $V_{555}$ observations both from the same epoch in 2006 -- see Table~\ref{table:obs}. The magnitudes are in the STMAG system. The location of the CV is marked with error bars.}\label{fig:cmd}
\end{figure}
 
Following the 1999 event recorded by \citet{sahu01}, the system was observed in a bright state about once per year up until 2004. \citet{piet05} estimated the outburst recurrence time for the system to be greater than 150~d based on their own observations as well as the 1999 event. Our observations are in agreement with this estimate. In nine months of monitoring we saw only one outburst of duration $\sim15$~d. The sampling was every second night for most of the  observing programme, but after discarding the images with the worst seeing, we are left with gaps in the data of up to 17~d. This means there is a possibility we may have missed another similar outburst; also, there is a stronger likelihood of any shorter outbursts of only a few days duration -- if present -- having gone undetected.

\subsection{Optical colours}\label{sec:colours}
In archival {\it HST}/WFPC2 data from several epochs covering the period 1994--2000, \cite{ack03} found that the object initially identified by \cite{sahu01} as a microlensing event was 0.5~mag brighter through a H$\alpha{}$ filter than main-sequence stars of the same magnitude. It was also a cluster member and coincident with a {\it ROSAT} source. This combination of factors led them to identify the object as the first known CV in the cluster (CV1). Interestingly, however, the broadband $(B_{439}-R_{675})$ and $(V_{606}-I_{814})$ colours place the star to the red side of the main sequence, in the region generally associated with detached main-sequence binaries. By contrast, CVs typically appear bluer than main sequence stars owing to the presence of a hot accretion disc. 
 
As discussed in Section~\ref{sec:hstdata}, in order to further investigate the broad-band colours of CV1, we analysed
archival {\it HST} images from 2006 taken in near-UV, {\it U} and {\it
V} filters. In the STMAG photometric system, we find
$V_{555}=19.25\pm0.05$ for the CV, with colours
$(U_{336}-V_{555})_{0}=0.19\pm0.12$~mag and
$(nUV_{255}-U_{336})_{0}=0.20\pm0.25$~mag, assuming a foreground
reddening of $E(B-V)=0.34$ (with an error of ten per cent) in the
direction of M22 \citep{harris96}. (We have chosen the STMAG system
because it is better calibrated than the synthetic WFPC2 system at UV
wavelengths.) Converting to the Johnson system (see
Section~\ref{sec:hstdata}), our measurements show that at $(U-V)_{0} =
0.13\pm0.13$~mag CV1 does not appear as UV-bright as field CVs in
quiescence, which typically have $(U-V)$ colours in the range $-0.1$
to $-0.7$~mag \cite[e.g.][]{bruch94}. This result is not wholly unexpected if we consider that the red colours found by \citet{ack03} suggest that the secondary contributes significantly to the red optical light -- perhaps down even as far as the {\it V}-band. If this is indeed the case, the optical emission of the donor would rival the UV emission from the disc, primary or hotspot, resulting in a less pronounced UV colour excess for the system than seen in typical field CVs.
In any case, we still see that in the
$U_{336}$ versus ($U_{336}-V_{555}$) CMD CV1 does have a UV excess (of
$\sim0.4$~mag) compared to the globular cluster main sequence -- as
would be expected for CVs (see Fig.~\ref{fig:cmd}).

Using the $(U-V)_{0}$ colour to constrain the temperature, at a distance to M22 of $3.2\pm0.3$~kpc \citep{harris96}, we calculate $L_{\mathrm{UV}} \sim 2.2\times 10^{32}$~erg~s$^{-1}$, over the UV wavelength range 2400--3400~\AA, for a blackbody spectrum of $\sim9000$~K. The uncertainty in this value is of the order of 40 per cent.

\subsection{X-ray characteristics}
\begin{figure}
\includegraphics[width=0.7\linewidth,angle=270]{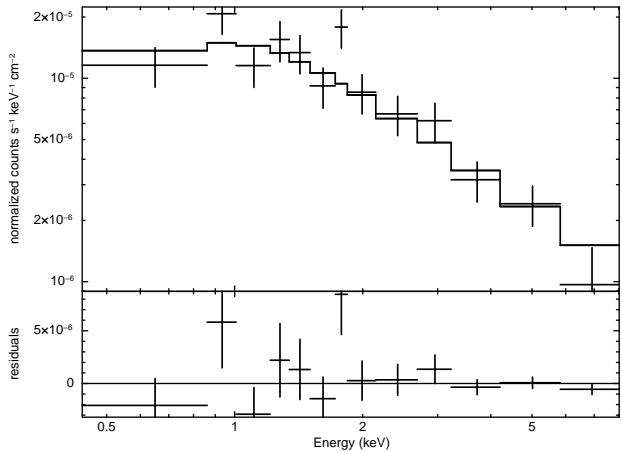}
 \caption{The Chandra ACIS-S spectrum of M22 CV1 fitted with a power law model (\xspec{} model parameters: \texttt{wabs(po)}, $N_H\sim 2.2\times10^{21}$~cm$^{-2}$, $\Gamma \sim 1.5$, $\chi^2_{\mathrm{reduced}} = 1.12$). The residuals of the fit are shown in the lower panel.}\label{fig:m22xspec}
\end{figure}

The X-ray spectrum (Fig.~\ref{fig:m22xspec}) is well fitted by a power law as well as other simple models such as a thermal bremsstrahlung. The parameters of the various fits are given in Table~\ref{table:xray}: the photon index we find for the power-law fit ($\Gamma=1.51\pm0.11$) and the temperature of the Bremsstrahlung fit (kT$_{\mathrm{Brems}}=10.00\pm4.07$~keV) are consistent with those found by \citet{webb04} ($\Gamma=1.70\pm0.13$; kT$_{\mathrm{Brems}}=9.69\pm3.11$~keV). \citeauthor{webb04} also found an absorption line at 1~keV in their {\it XMM-Newton} spectrum of the object. We checked for the presence of this line by adding a Gaussian absorption line at 1~keV. In contrast to the results of \citeauthor{webb04}, this did not improve the fit. Considering the relatively poor signal-to-noise of our data, we also added a simulated absorption line at the same estimated normalization as that of \citeauthor{webb04} to examine the effect on the fit. The addition significantly worsened the power law fit, suggesting that no such line is present in our data. 

\begin{table*}
 \centering
 \begin{minipage}{140mm}
  \caption{Parameters of the best-fitting spectral models to the {\it Chandra} ACIS-S data for CV1. The flux given is the unabsorbed flux in the 0.3--10~keV energy range. The hydrogen column was frozen to the value for the cluster \citep[$2.2\times10^{21}$~cm$^{-2}$, ][]{webb04}}\label{table:xray}
  \begin{tabular}{@{}ccccccc@{}}  
  \hline
  N$_H$ & Model & kT (keV) & Photon & $\chi^2$ & d.o.f. & Flux \\     
  $\times10^{21}$~cm$^{-2}$  & & & Index & & & $\times10^{-14}\, $ergs$\, $cm$^{-2}$~s$^{-1}$  \\

 \hline

  2.2		& PL 								& - 			& 1.51$\pm$0.11 & 1.12 	& 11 	& 21.5  \\
  2.2		& Brems.							& 10.00$\pm$4.07	& -		& 0.98	& 11	& 19.7	\\
  2.2           & RS \footnote{Raymond Smith}                                                           & 11.54$\pm$4.80        & -             & 1.22  & 11    & 21.2      \\
  2.2		& PL+Gau.\footnote{line centre = 1.00$\pm$0.05, $\sigma= 0.1$, normalized to drop to 0.6 of continuum}	& -	& 1.61$\pm$0.11 & 1.81	& 10	&	\\

 \hline
\end{tabular}
\end{minipage}
\end{table*}

The X-ray spectral-hardness ratio was calculated for CV1 from the Chandra data in the same manner as \citet{heinke05} using the same equation they used for the plot in their fig. 9,
\begin{equation}
\label{eq:heinke}
X_{\mathrm{colour}} = 2.5\, \mathrm{log}\frac{F_{X[0.5\text{--}1.5~\mathrm{keV}]}}{F_{X[1.5\text{--}6~\mathrm{keV}]}}.
\end{equation}
The hardness of its X-ray colour, $X_{\mathrm{colour}}=-0.39$, combined with the X-ray luminosity we calculate in the 0.5--6~keV band, $L_{\mathrm{X[0.5\text{--}6~\mathrm{keV}]}}= 1.4\times10^{32}~$ergs~s$^{-1}$, places CV1 in the region of the plot of \citeauthor{heinke05} occupied by the 47 Tuc CVs. 
By contrast, quiescent LMXBs have significantly softer X-ray colors \citep[see, e.g., fig.~10 of][]{heinke05} even though their X-ray luminosities, though usually higher, range down to a similar $\sim10^{32}$~ergs~s$^{-1}$. We conclude that the identification of this source as a cataclysmic variable is correct.

\section{M5 V101}

\subsection{Spectrum of M5 V101}
\begin{figure}
\vspace{3mm}
\centering
\includegraphics[width=0.7\linewidth,angle=270]{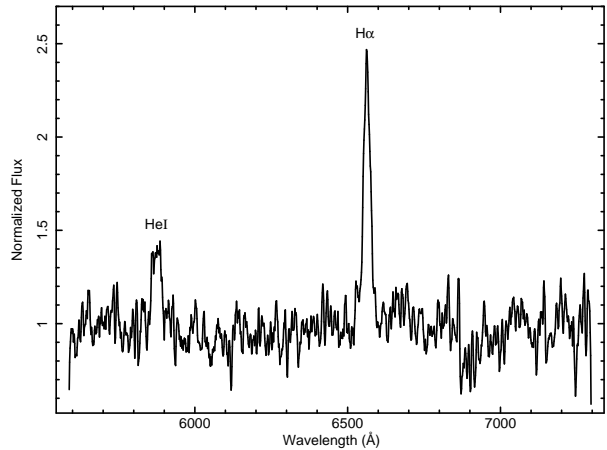}
\caption{The red summed WHT ISIS spectrum of M5 V101 in quiescence.}
\label{fig:redspec}
\vspace{3mm}
\end{figure}

\begin{figure}
\vspace{3mm}
\centering
\includegraphics[width=0.7\linewidth,angle=270]{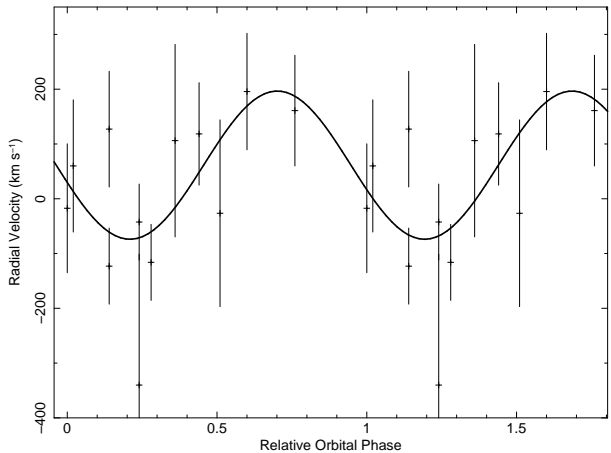}
\caption{The radial velocities from the H$\alpha{}$ emission-line fits folded on the orbital period.}
\label{fig:rv}
\vspace{3mm}
\end{figure}

H$\alpha{}$ $\lambda 6563$~\AA{} appears as the strongest emission line in the red spectra, with an equivalent width of $50\pm5$~\AA{} in the summed spectrum (Fig.~\ref{fig:redspec}). He\,{\sevensize I} $\lambda 5876$~\AA{} also appears in emission with an equivalent width of $8.5\pm2$~\AA. The absorption appearing at 6867--6884~\AA{} is the Fraunhofer B band caused by molecular O$_{2}$ in the atmosphere. The weak absorption feature at 7118~\AA{} may be associated with the secondary star, but the origin of other apparent features, such as at 6118\, \AA{} and 7245\, \AA, cannot be identified (they do not seem to be telluric in nature, for example).  
The H$\alpha{}$ emission line was fitted with a Gaussian profile in the individual spectra and the resulting radial velocity curve -- phased on the 5.796-h orbital period \citep{neill02} -- is shown in Fig.~\ref{fig:rv}. Despite the poor signal-to-noise, we find that the $\gamma{}$ velocity from the fit, $57\pm35$~km~s$^{-1}$, is consistent with the cluster velocity of $52.6\pm0.4$~km~s$^{-1}$ for M5 \citep{harris96}.

\begin{figure}
\includegraphics[width=0.7\linewidth,angle=270]{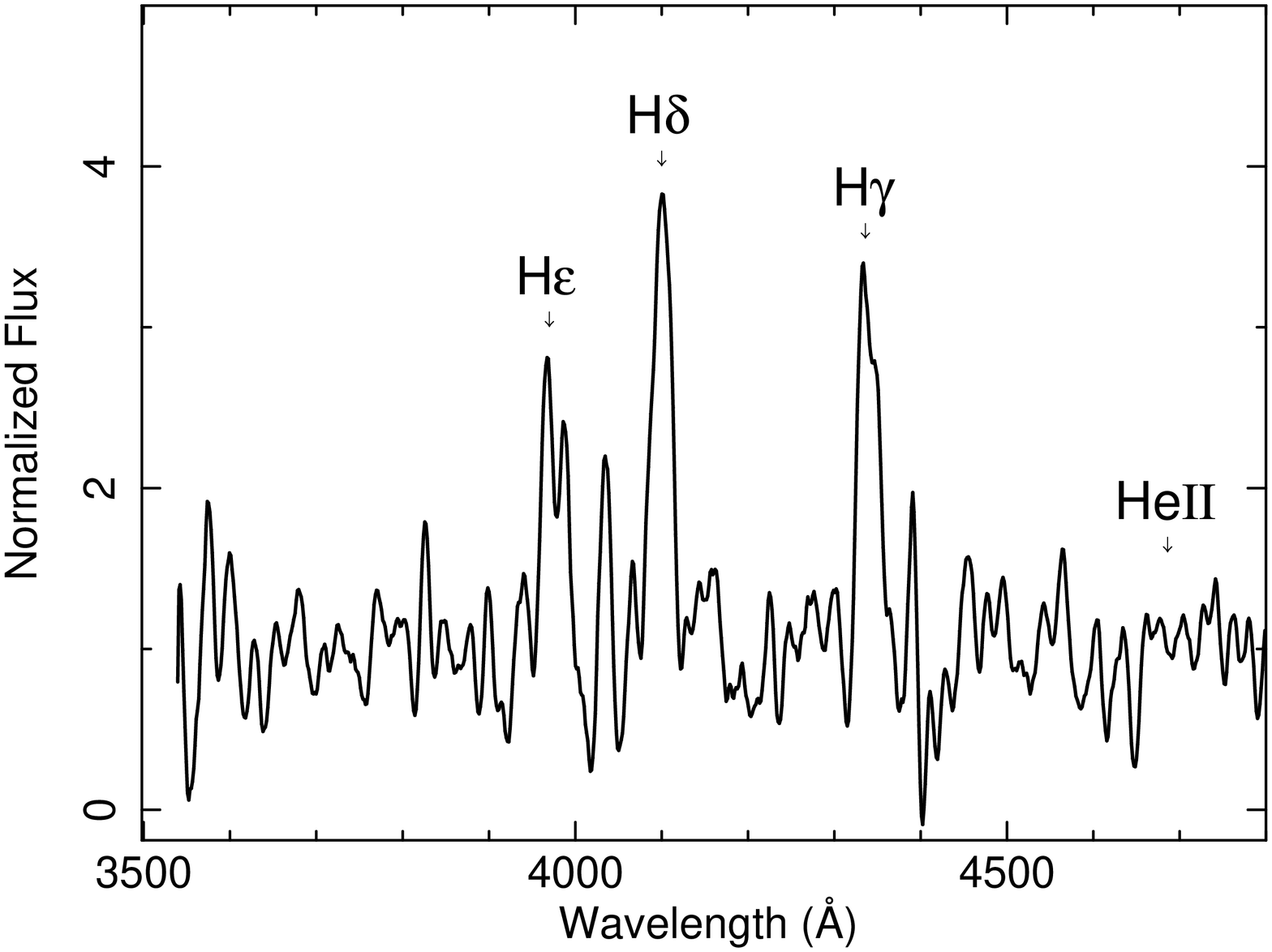}
\caption{The blue summed WHT ISIS spectrum of M5 V101 in quiescence.}
\label{fig:bluespec}
\end{figure}

The summed blue spectrum (Fig.~\ref{fig:bluespec}) displays prominent Balmer emission but there is no evidence for any He\,{\sevensize II} emission.

A list of all the emission line measurements is given in Table~\ref{table:eqw}.

\begin{table*}
 \centering
 \begin{minipage}{140mm}
  \caption{Characteristics of emission lines present in M5V101 spectrum}\label{table:eqw}
  \begin{tabular}{@{}lccc@{}}  
  \hline
  Line & Wavelength & Equivalent Width  & Velocity FWHM \\     
  & (\AA) & (\AA) & (km~s$^{-1}$) \\
 \hline
H$\alpha{}$ & 6563 & 50$\pm3$ & 1100$\pm150$ \\
He\, {\sevensize I} & 5876 & 22$\pm5$ & 1800$\pm100$ \\
H$\gamma{}$ & 4340 &  90$\pm30$ & 1900$\pm200$  \\
H$\delta{}$ & 4102 &  80$\pm20$ & 1700$\pm150$ \\
H$\epsilon{}$ & 3970 &  60$\pm30$ & 2300$\pm300$ \\
 \hline
\end{tabular}
\end{minipage}
\end{table*}

\subsection{Photometry}\label{sec:model}

\citet{neill02} found $P_{\mathrm{orb}} = 5.796\pm0.036$~h for M5 V101 from {\it I}-band photometry. They saw two peaks in the orbital modulation in the redder {\it I}-band light curve and only one peak per orbit in the {\it V}-band, leading them to interpret one of the modulations as being associated with the secondary, which, at K5--M0 (which they estimated from their $(V-I)$ colour, $\simeq 2.2$), would be much fainter in the {\it V}-band. 

\begin{figure}
\centering
\begin{tabular}{c}
\includegraphics[width=0.9\linewidth,angle=0]{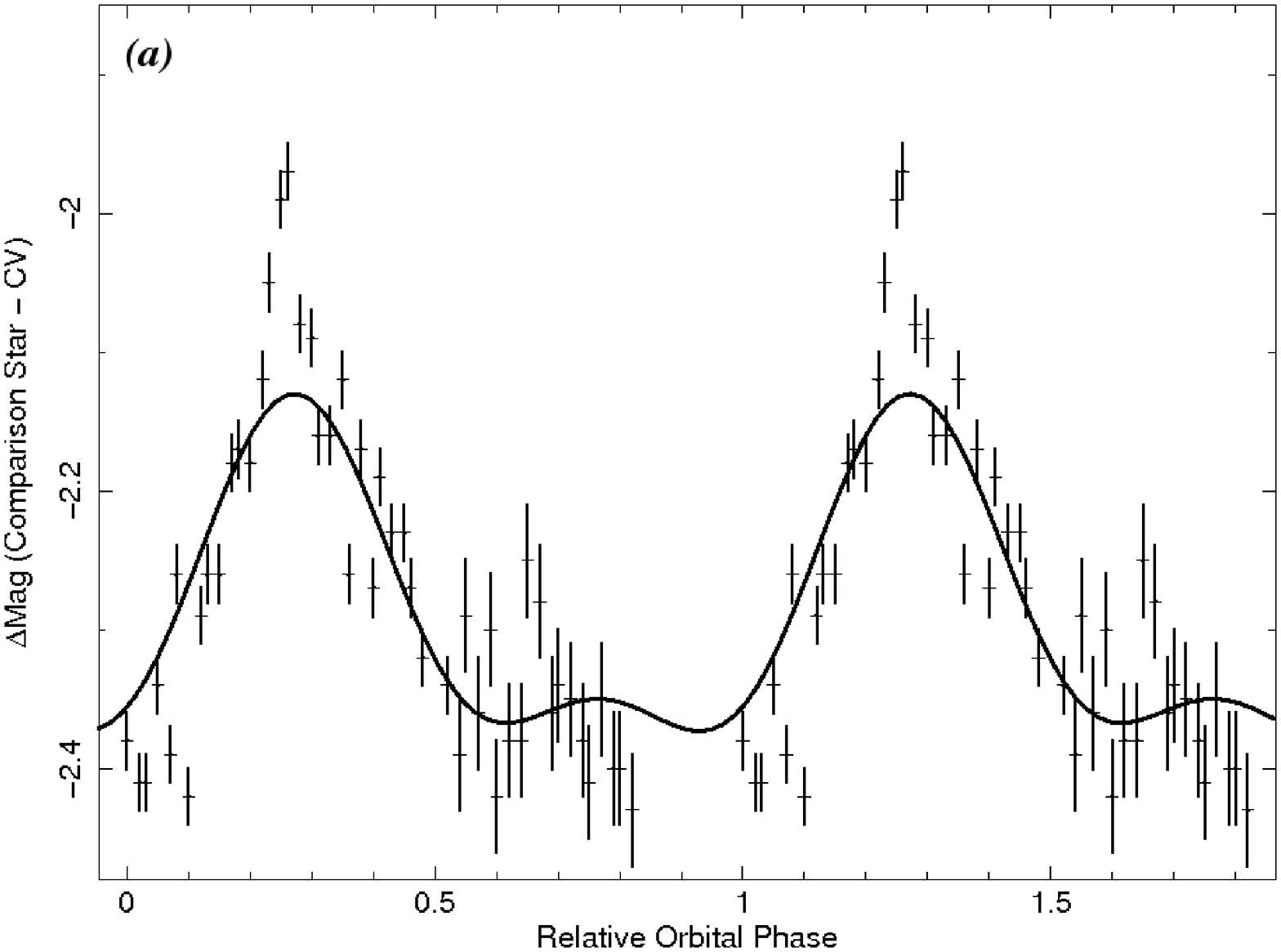}
	\\
\includegraphics[width=0.9\linewidth,angle=0]{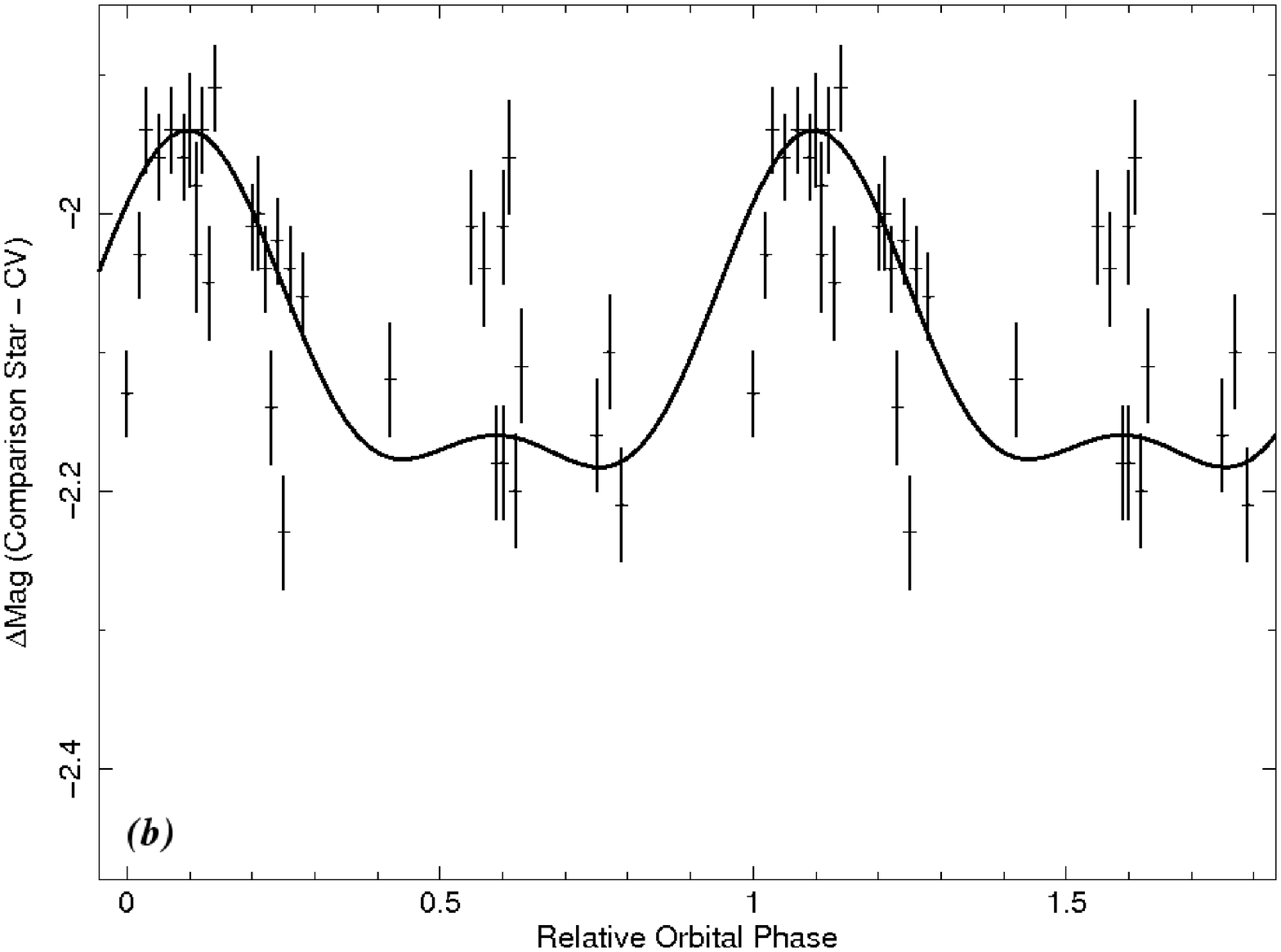}
	\\
\includegraphics[width=0.9\linewidth,angle=0]{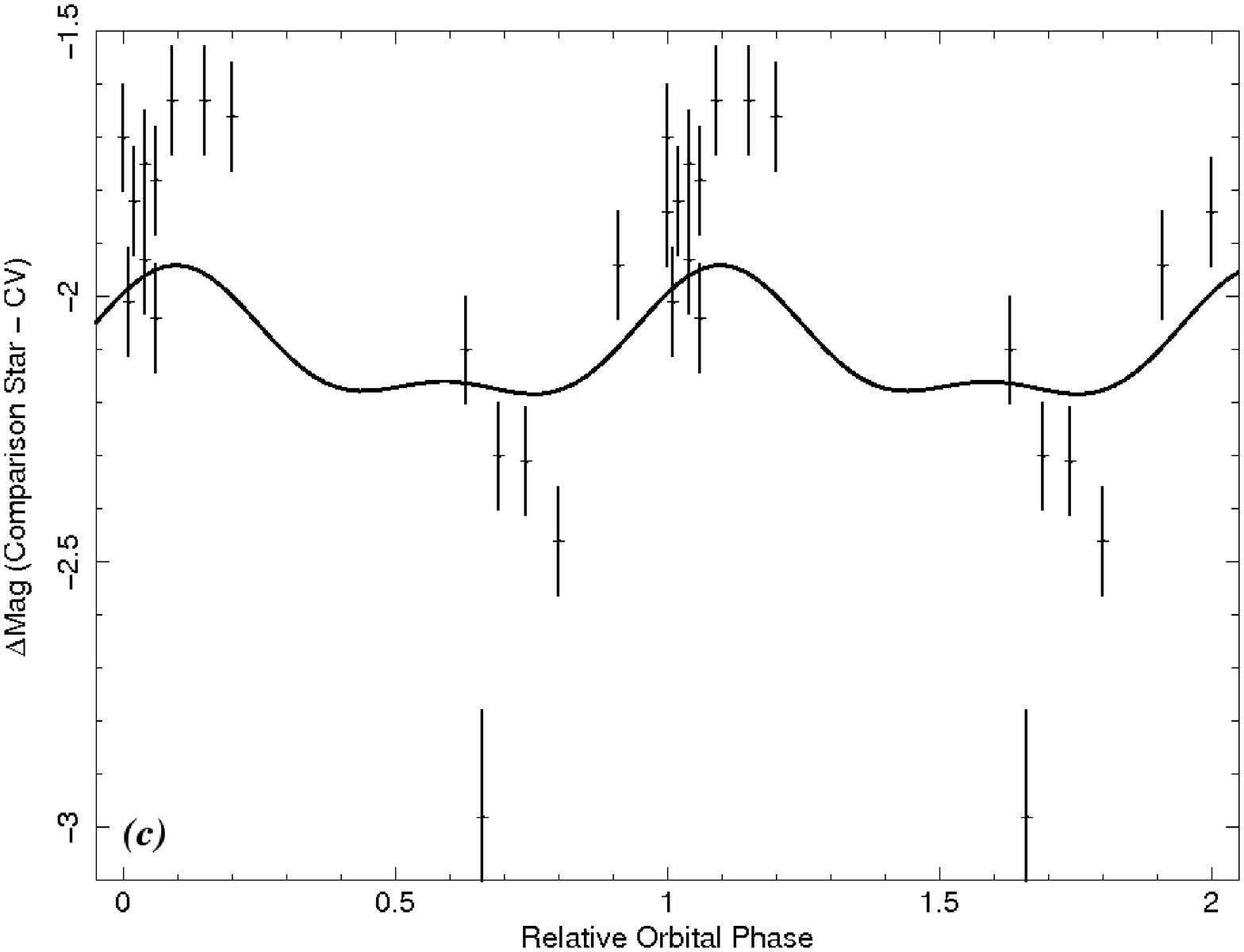}
\vspace{5pt}
\end{tabular}
\caption{The M5 V101 light curves folded on the orbital period, $P_{\mathrm{orb}}=5.796\pm0.036$~h: (a) 1990 April INT {\it R}-band data; (b) 1990 June INT {\it R}-band data; (c) 2009 July MDM {\it r}-band data. We have superimposed the model of \citet{neill02} (see Section~\ref{sec:model}) which they fit to their \textit{V}-band data, using their model parameters, on our folded light curves.}\label{fig:lcint}
\end{figure}

Our INT {\it R\/}-band light curve for the night of 1990 April 1 (Fig.~\ref{fig:lcint}~(a)) exhibits a 0.3-mag modulation at the orbital period of \cite{neill02}. The modulation does not appear as strongly in the June data (Fig.~\ref{fig:lcint}~(b)). Some flickering of the object was observed at this time. We superimposed the same model (Equation~\ref{eq:model}) used by \citeauthor{neill02},
\begin{equation} y = a + b sin(c\phi + d) + e sin(f\phi+g) \label{eq:model};
\end{equation}
 on our folded light curves for comparison: the resulting plots are shown in Fig.~\ref{fig:lcint}. The main peak we observe in the orbital light curve is probably due to the hotspot (the region where the accretion stream from the secondary impacts the accretion disc) rotating into view while the absence of any eclipse suggests the inclination is not high (\citeauthor{neill02} estimated the inclination to be between 50--60$\degr$). 
We find reasonable agreement with the results of \citeauthor{neill02}: our {\it R}-band data show an intermediate situation between what they observed in the {\it V} and {\it I} bands, with a second but fainter peak visible in the orbital modulation. 

The CV was on average $0.2\pm0.1$~mag brighter during the MDM observations than during the INT observations. Thus, it might be expected that the increased luminosity of the accretion disc would dilute the contribution from the secondary, removing the associated second peak -- if such is its source -- from the orbital modulation. Unfortunately, the phase coverage in our MDM {\it r}-band folded light curve (Fig.~\ref{fig:lcint}~(c)) is insufficient to enable us to discern the morphology of the orbital modulation.

\subsection{X-ray characteristics}
The X-ray luminosity of the source in the 0.5--2.5~keV band was calculated by \cite{hakala97} to be $L_{\mathrm{X}} \simeq 1.1 \times 10^{32}$~erg~s$^{-1}$. We extrapolate the flux they measured to find the flux in the 0.5--6~keV band by fitting with a 3-keV bremsstrahlung model in WebPIMMS\footnote{http://heasarc.nasa.gov/Tools/w3pimms.html}, finding $F_{\mathrm{X}} \simeq 2.4 \times 10^{-14}$~erg~s$^{-1}$~cm$^{-2}$ (unabsorbed). We calculate the UV and optical flux using the quiescent {\it V}-band magnitude of $20.27\pm0.04$~mag from \citet{kaluz99} with the formula, log$F_{\mathrm{uv+opt}}=-0.4m_{\mathrm{V}} - 4.32$, from \citet*{teese96}, to find $F_{\mathrm{uv+opt}} \simeq 3.7 \times 10^{-13}$~erg~s$^{-1}$~cm$^{-2}$. This leads to an X-ray-to-UV/optical flux ratio of $F_{\mathrm{X}}$/$F_{\mathrm{uv + opt}}\sim 0.06$. Previously, the system was seen as faint as $V=22.5$ \citep{kukarkin70}, which would lead to an $F_{\mathrm{X}}$/$F_{\mathrm{uv + opt}}$ ratio a factor of ten higher, but there is no evidence for the system being this faint during the epoch from which we take the measurements that we use in our calculation. Nevertheless, since we cannot be sure that the system was not fainter at the time the X-ray flux was measured, we consider the $F_{\mathrm{X}}$/$F_{\mathrm{uv + opt}}$ ratio estimated above to be a lower limit.

\section{Discussion}

\subsection{M22 CV1}

\subsubsection{X-ray properties: CV or LMXB? Magnetic or non-magnetic?}

Previously \citet{webb04} found $\Gamma=1.70\pm0.13$ for a power law
fit to archival \textit{XMM-Newton} X-ray spectral data for the
source; in addition they found an even softer power law photon index
if they added a Gaussian absorption line at 1~keV, which improved
their fit. They interpreted this absorption line as possible electron
cyclotron resonance in a neutron star pulsar. The lack of any evidence
for such an absorption line in our data weakens the case for a neutron
star pulsar (or LMXB) identification for CV1. The hardness of the
X-ray colour of the object, $X_{\mathrm{colour}}=-0.39$, is also more
consistent with a CV than a quiescent LMXB. Thus, based on these
X-ray properties of the system, it appears much more like a CV than a
qLMXB. 

We also calculated the X-ray luminosity in the {\it Chandra} ACIS
0.3--10~keV band, $L_{\mathrm{X}}\simeq2.2\times10^{32}$~erg~s$^{-1}$,
and found it is at the high end of the range of X-ray luminosities for
non-magnetic dwarf novae, but is typical for the magnetic IPs. Our
best-fitting power-law spectrum is relatively hard, with photon index
$\Gamma=1.51\pm0.11$. \citet{heinke08} investigated the relationship
between CV subtype and X-ray spectrum and luminosity for a sample of
CVs comprising confirmed magnetic CVs (IPs and polars), quiescent DN
and nova-like CVs (NL) and globular cluster CVs. They found the
distribution of photon index differed significantly between magnetic
CVs, which are harder X-ray sources (mean $\Gamma=1.22, \sigma=0.33)$,
and non-magnetic CVs (mean $\Gamma=1.97, \sigma=0.20)$, but was fairly
consistent within either group. The cluster sources were found to
comprise $\sim 40$ per cent magnetic systems, based on the photon
index distribution. Our measured photon index of $\Gamma=1.51\pm0.11$
places CV1 on the borderline between the magnetic and non-magnetic
distributions. However, the combination of our result with the photon
index result of \citet{webb04}, which, although consistent with our
value, ranges to softer values, tips the balance slightly in favour of a
non-magnetic CV classification for the object.

\subsubsection{Outburst characteristics}\label{sec:outburst} 
In the optical, the characteristics of the $\sim$15-d 2004 May
outburst of M22 CV1 covered by our observations are consistent with
previous reports of $\sim$15--20~d, 2--3~mag eruptions of the system
by \citet{sahu01}, \cite{bond05} and \cite{piet05}. These outburst
characteristics are also consistent with normal outbursts of DN, which
typically have durations from 2--20~d \citep{warner95}, where this
duration is also correlated with the recurrence time for
outbursts. \cite{piet05} estimated the outburst recurrence time for
CV1 to be greater than 150~d -- our observations are consistent with
this limit.

The typical duration of outbursts of CV1, at 15--20~d, would
tend to favour a longer orbital period estimate of the order of 10~h
or more \citep*[e.g.][]{ak02}. In addition, the outburst properties of
CV1 are not unlike those of non-magnetic dwarf novae of longer orbital
period, such as BV Cen, which undergoes outbursts with an amplitude of
$\sim3$~mag and a 30-d duration separated by 150~d (although these
tend to be more symmetrical in rise and decay than those of CV1).

If \emph{magnetic}, CV1 cannot be a polar because their absent discs
preclude outbursts. On the other hand, intermediate polars (IPs) --
with their partially truncated inner discs -- do exhibit
outbursts. However, the outburst characteristics of CV1 are not consistent with any
particular example of the known IPs. GK Per, with its infrequent
outbursts which persist for two months, remains an anomalous
system. Other IP systems such as EX Hya exhibit a similar outburst
amplitude to CV1, $\sim3.5$~mag, but these outbursts last only 2--3~d
and have a longer recurrence time of $\sim2$~yr \citep{hell89}, while
IPs such as TV Col and V1223 Sgr show even shorter duration
low-amplitude outbursts that last only $\sim0.5$~d. 
Thus, on the basis of its outburst properties, CV1 does not resemble a magnetic system.
We can also eliminate the \emph{SU UMa} class based on the
morphology of the outburst light curve: even though our sampling is
too sparse to detect superhumps (the modulation which is the defining
characteristic of the SU UMa class) in the ouburst light curve, we find
no evidence for the extended, sloping
plateau of brightness associated with superoutbursts. Similarly,
\citet{piet05} found no evidence for the presence of superhumps in the
2000 August outburst light curve. This leaves the longer-period
\emph{U Gem}-type DN as the group of CVs with the most similar
properties to CV1 (e.g. BV Cen).

\subsubsection{Optical colours}
\citet{echev84} show that for a sample of field DN, the $(B-V)$ and
$(U-B)$ colours are well correlated with orbital period, with those
systems having $P_{\mathrm{orb}}<5\frac{1}{2}$~h being significantly
bluer than those with $P_{\mathrm{orb}}>7$~h. Conversely, for the same
sample, the $(V-R)$ and $(R-I)$ colours do not appear to be well
correlated with $P_{\mathrm{orb}}$. Based on its $(U-V)_0$ colour of
$0.13\pm0.13$~mag, CV1 appears to be consistent with having
$P_{\mathrm{orb}}> 7$~h. 
The blackbody temperature of $\sim 9000$~K inferred from the
$(U-V)_0$ colour is consistent with emission from the accretion
disc. 

\citet{ack03} found that the ($B_{439}-R_{675}$) and ($V_{606}-I_{814}$) colours of CV1 are unusually red for a CV, about 0.2~mag and 0.1~mag redward of the main sequence, respectively. 
A possible explanation for these red colours, previously mentioned by \citeauthor{ack03}, is that a secondary larger than a normal main-sequence star dominates the optical emission of the system. 
\citet{baraffe00} show that for longer orbital period ($P_{\mathrm{orb}}\gtrsim6$~h) CVs, the degree of nuclear evolution of the secondary is crucial in setting its spectral type, with the most evolved donors having spectral types significantly later than their main-sequence analogues. 
Some degree of nuclear evolution of the donor off the main sequence could explain the location of CV1 in the CMDs of \citeauthor{ack03}. 
An enhanced contribution to the {\it V}-band from such a secondary might also explain the relatively low UV excess inferred from the ($U-V$) colour (see Section~\ref{sec:colours}).
Although an established example of an evolved globular cluster CV secondary may be found in the dwarf nova AKO~9 in 47 Tuc (see the CMDs of \citealt{albrow01}; \citealt{edmonds03a} and the calculations of \citealt{knigge03}), the ($V-I$) colours of many other globular cluster CVs \citep[e.g.][]{cool98,edmonds03a} suggest a lesser, if any, degree of nuclear evolution of the donors in these systems.

We note that the location of CV1 on the ($B_{439}-R_{675}$) and ($V_{606}-I_{814}$) CMDs of \citeauthor{ack03},
 together with its UV and X-ray colours, could also be explained
if CV1 were part of a hierarchical triple system in which the third
star were a main-sequence star of mass comparable to the secondary
star in CV1. It would be of interest to determine whether such a
system could survive in the core of M22. 
Alternatively, the ($B_{439}-R_{675}$) and
($V_{606}-I_{814}$) colours could be accounted for by the presence of a line-of-sight
main-sequence star. The relatively low density of stars in the
vicinity of CV1 in the \textit{HST} image make this unlikely, but it cannot be ruled out.
 
\subsubsection{X-ray to optical/UV flux ratio: the CV subclass of CV1}
A useful discriminant of the subtype of a CV is the ratio of X-ray to
ultra-violet and/or optical flux. 

\citet{verbunt97} demonstrated that the different
classes of CV can be 
subdivided as follows: SU UMa and U
Gem-type CVs, $ F_{\mathrm{X}} $/$ F_{\mathrm{uv + opt}} \sim0.1$; $
F_{\mathrm{X}} $/$ F_{\mathrm{uv + opt}} \sim0.01$ for Z Cam-type CVs;
finally, UX UMa-type (or \textit{novalike}) CVs have the lowest ratios
at $ F_{\mathrm{X}} $/$ F_{\mathrm{uv + opt}} \sim 0.001$.

Calculating the ratio of X-ray to UV plus optical flux in an
equivalent manner for CV1, we find $F_{\mathrm{X}}$/$F_{\mathrm{uv +
opt}} \sim $0.07 (where the discrepancy with the result of
\citet{ack03} arises from the inclusion of the UV flux
here). This value again suggests a U Gem classification for CV1.

\subsubsection{X-ray to optical/UV flux ratio: constraining the 
orbital period of CV1}

The orbital period distribution of CVs typically ranges from tens of
minutes up to about 15~h. The value of $P_{\mathrm{orb}}$ for a CV can
reveal much about the parameters of the system
\citep[e.g.][]{warner95}. 

Whilst it is challenging to directly measure the orbital period for a
system as faint and crowded as CV1 in M22, we can none the less attempt
to constrain it by comparing the optical, UV and X-ray properties of
the system to those of Galactic CVs with known orbital periods. For
example, the absolute magnitude at minimum of the source,
$M_{\mathrm{V}} \sim 5.4$~mag \citep{ack03}, suggests an orbital
period of $P_{\mathrm{orb}} = 13\pm3$~h (see equation 18 of
\citet{warner87}).

Furthermore, because of the increased UV luminosity of CVs at longer
orbital periods due to their correspondingly higher accretion rates
\citep{teese96}, and the lack of any clear correlation between X-ray
luminosity and orbital period, it would seem reasonable to expect a
correlation between $ F_{\mathrm{X}} $/$ F_{\mathrm{uv + opt}}$ and
orbital period for non-magnetic CVs. In fact, \citeauthor{teese96}
 have found evidence for such a correlation: their
fig. 6 shows that for a sample of non-magnetic field CVs, with
increasing orbital period there is a trend of decreasing
$F_{\mathrm{X}}$/$F_{\mathrm{uv + opt}}$.

Interestingly however, our $F_{\mathrm{X}}$/$F_{\mathrm{uv + opt}}$
value for CV1 places it in the region of the \citeauthor{teese96} plot with
$P_{\mathrm{orb}}\lesssim 2$~h, in contradiction to our previous
orbital-period constraints. We discuss below two ways in which this could be
resolved:

\begin{enumerate}
\renewcommand{\theenumi}{(\arabic{enumi})}
\item Taking the contradiction with the van~Teesling results at face
value, one way to accommodate such a short orbital period would be to
invoke the presence of a cooler (redder) line-of-sight
interloper or a companion in a triple system. Then the intrinsically fainter, bluer CV would be more
consistent with a shorter $P_{\mathrm{orb}}$. In order to revise the
$P_{\mathrm{orb}}$ estimate using the relationship from
\citet{warner87} down to $2\pm1$~h, an absolute magnitude at minimum
as faint as $\sim$9~mag would be required. However, given the measured
H$\alpha$ excess of 0.54~mag \citep{ack03}, a fainter CV would have a
larger H$\alpha$ excess translating to an unfeasibly large equivalent
width of $\sim 330$~\AA{} for the emission line, making this an
unlikely explanation.

\item A more likely explanation comes from a closer examination of the
data used by \citeauthor{teese96}: none of these
systems are, in fact, U Gem types (the solitary datum of this class, EI
UMa, has since been reclassified as an IP \citep{ramsay08}). 

\end{enumerate}

Hence, we conclude that our classification of CV1 as a dwarf nova of U Gem type remains secure.

\subsection{M5 V101}

Strong He\,{\sevensize II} $\lambda 4686$~\AA{} line emission is characteristic of magnetic CVs 
\citep[in polars, it is comparable in strength to H$\beta$;][]{warner95}. 
In IPs, for example, this high-excitation line is most likely powered by X-ray heating of the
regions of the accretion curtains close to the magnetic poles, interior to the inner
edge of the magnetically truncated accretion disc \citep{saito10}.
The strong He\,{\sevensize I}, H$\alpha{}$ 
and Balmer emission (see Figs.~\ref{fig:redspec} and \ref{fig:bluespec}) of V101 are
characteristic of CVs in quiescence, while the lack of any obvious
He\,{\sevensize II} emission -- an upper limit of $6$~\AA{} can be
placed on the equivalent width of any He\,{\sevensize II} $\lambda
4686$~\AA{} emission in the spectrum
-- suggests that the system is unlikely to be magnetic.

The enhanced contribution of the accretion disc to the luminosity of
the system at shorter wavelengths means we would expect the absorption
lines of the secondary star to be more apparent at longer
wavelengths. The only possible absorption feature from the secondary
which can be identified in our summed spectrum is a weak feature at
7118~\AA{}, although we cannot be certain this is not an artefact.

Our phased optical light curves in the {\it R} and {\it r} bands show
varying degrees of modulation at the orbital period of
\cite{neill02}. The main peak in the light curve may be due to the
hotspot \citep[see e.g.][]{wood86}.
Comparing our $F_{\mathrm{X}}$/$F_{\mathrm{uv + opt}}\geq0.06$ to the
trend in fig. 6 of \citet{teese96}, as we did for M22 CV1, again we
find that for our result, their trend would predict a shorter orbital period of up
to a maximum of around three hours, compared to the measured
$P_{\mathrm{orb}} = 5.796\pm0.036$~h \citep{neill02}. This discrepancy
is not as pronounced as what we found for the M22 CV, although in that case we did
not have a measured $P_{\mathrm{orb}}$ to compare with. However, again
we note the lack of U Gem-type systems in the \citet{teese96} sample.

\section{Summary}

For M22 CV1, the X-ray properties confirm a CV rather than an LMXB
nature. The high $L_{\mathrm{X}}$ and X-ray spectral hardness are indicative
of either a quiescent DN or an IP. The outburst characteristics are
consistent with a normal long $P_{\mathrm{orb}}$ ($\gtrsim 10$~h)
dwarf nova of U Gem type, but cannot be very easily reconciled with
any particular example of the known IPs. We find the expected UV excess
compared to the cluster main sequence in $U_{336}$ versus ($U_{336}-V_{555}$), and that the
($U-V$) colour is consistent with a longer orbital period. 

For the M5 CV, V101, our optical spectra are typical for a quiescent
dwarf nova. The lack of any significant He\,{\sevensize II} emission
suggests the system is not magnetic. The modulation we observe in the
{\it R}-band light curve is consistent with the orbital period of
$5.796\pm0.036$~h found by \citet{neill02}. As this remains the  
globular cluster CV most amenable to detailed study from the ground in
quiescence, it will benefit from future higher S/N phase-resolved optical spectroscopy.

\section*{Acknowledgments}
We would like to thank the referee, Christian Knigge, for helpful comments and suggestions.
This research has made use of data obtained from the {\it Chandra Data Archive} and software provided by the {\it Chandra X-ray Center\/} ({\it CXC}) in the application package {\sevensize CIAO}. This research has also made use of the SIMBAD database, operated at CDS, Strasbourg, France.

\label{lastpage}

\end{document}